\documentclass[utf8
               ]{jetp}
\twocolumn
\begin{document}

\title{Post-Newtonian limit of hybrid metric-Palatini f(R)-gravity}

\author{P.~I.}{Dyadina}
\email{guldur.anwo@gmail.com}
\affiliation{Sternberg Astronomical Institute, Lomonosov Moscow State University, Universitetsky Prospekt, 13, Moscow, Russia}

\author{S.~P.}{Labazova}
\email{sp.labazova@physics.msu.ru}
\affiliation{Department of Astrophysics and Stellar Astronomy, Lomonosov Moscow State University, Leninskie Gory, 1/2, Moscow, Russia}
\author{S.~O.}{Alexeyev}
\email{alexeyev@sai.msu.ru}
\affiliation{Sternberg Astronomical Institute, Lomonosov Moscow State University, Universitetsky Prospekt, 13, Moscow, Russia}
\affiliation{Department of Quantum Theory and High Energy Physics, Lomonosov Moscow State University, Leninskie Gory, 1/2, Moscow, Russia}

\rtitle{Post-Newtonian limit of hybrid metric-Palatini f(R)-gravity}
\rauthor{P.~I.~Dyadina, S.~P.~Labazova, S.~O.~Alexeyev}

\abstract
{Using the latest most accurate values of post-Newtonian parameters $\gamma$ and $\beta$  obtained by MESSENGER we impose restrictions  on the recently proposed hybrid f(R)-gravity model in its scalar-tensor representation. We show that the presence of a light scalar field in this theory does not contradict the experimental data based not only on the $ \gamma $ parameter (as was shown earlier), but also on all other PPN parameters. The application of parameterized post-Newtonian formalism to  gravitational theories with massive fields is also discussed.
}

	\maketitle

	\section{Introduction}\label{sec1}
	
	Currently, the General Relativity (GR) is a worldwide accepted theory of gravity. During more than one century a lot of problems were solved within such approach. However, as the quality of observations increased, new phenomena appeared that could not be explained within the framework of GR. For example, in the late XXth century, the problem of accelerated Universe expansion was discovered but the nature of this phenomenon has still not clear~\cite{DE1,DE2,DE3}. Another puzzle of modern physics is manifestations of dark matter (DM) on scales of galaxies and galaxy clusters~\cite{DM1,DM2}. These issues can be studied in two ways: by introduction of new particles  or by changing the geometry of space-time. The second way leads to the appearance of modified gravitational models, which are based on changing of GR.
	
	Among different ways of GR expansion f(R)-gravity should be identified especially~\cite{fr1, fr2, odintsov9, odintsov10}.  The f(R)-gravity is the simplest extension of the Einstein-Hilbert action. This theory is based on  generalization of the gravitational part of the action as an arbitrary function of the Ricci scalar $R$. Such models have become widespread after f(R)-gravity was successfully applied in inflation theories~\cite{starobinsky}. The f(R)-gravity is attractive because the accelerated expansion of the Universe is natural consequence of the gravitational theory. In addition, f(R)-gravity is interesting as an alternative to the $\Lambda$CDM model, since it allows to simultaneously describe early-time inflation and late-time cosmic acceleration~\cite{odintsov1, odintsov2, odintsov3, odintsov4, odintsov5, odintsov6, odintsov7,  saez}. Moreover, f(R)-models can provide good agreement with observational data, being almost indistinguishable from $\Lambda$CDM~\cite{odintsov8}.
	
	There are two possible approaches to obtain field equations from those modified actions: the metric one and the Palatini one. In the metric approach  $g_{\mu\nu}$ is the only dynamical variable and  the action is varied with respect to it only. The Palatini method is based on the idea of considering the connection defining the Riemann curvature tensor to be a priori independent of the metric. Thus, variations with respect to the metric and the connection are performed independently. Besides, the Palatini method provides second order differential field equations, while in the metric approach these equations are of the fourth order~\cite{capo1,capo2}.
	
	However, some shortcomings of f(R)-theory are manifested both in the metric and in the Palatini approaches. One of the main problems of metric f(R)-gravity is difficulties with passing standard tests in the Solar System~\cite{chiba, olmo1, olmo2}. Nevertheless, a limited class of viable models in the metric approach exists and was studied in detail in the papers~\cite{odintsov4, odintsov7, odintsov8}. Most clearly, the features of f(R)-gravity are manifested in its scalar-tensor representation. In particular, its metric version can be interpreted as the Brans-Dicke scalar-tensor theory with the parameter  $\omega_{BD} = 0$ ($\omega_{BD}$ is Brans-Dicke parameter) and nontrivial potential $V(\phi)$. In order for the theory to satisfy the constraints imposed by laboratory experiments and observations in the Solar System, the scalar field should be massive, with an interaction range not exceeding a few millimeters. Such a scalar field obviously cannot influence cosmological picture~\cite{khoury, capo0}. Thus, metric f(R)-theories are viable only when the scalar field can somehow be `` hidden '' in local experiments, while on cosmological scales it behaves like a long-range field.  This feature is achieved through the chameleon mechanism~\cite{khoury, khoury1, hu, odintsov3}.
	
	On the other hand, the Palatini $f(\mathfrak{R})$-model can be represented as the scalar-tensor Brans-Dicke theory with $\omega_{BD} =-3/2$ and the same potential $V(\phi)$ as in the metric approach. The theory is characterized by the presence of a non-dynamical scalar field. Suchl nature of the scalar implies that in vacuum Palatini $f(\mathfrak{R})$-model turns into GR with an effective cosmological constant $\Lambda_{eff}$. This property allows to describe the late-time cosmic acceleration, if $\Lambda_ {eff}$ is small enough. Despite this attractive property, all Palatini $f(\mathfrak{R})$-models with a small $\Lambda_{eff}$ that have been studied so far lead to microscopic matter instabilities and to unacceptable features in the evolution patterns of cosmological perturbations~\cite{koivisto1, koivisto2}.
	
	Recently, a theory constructed as a superposition of the Einstein-Hilbert metric Lagrangian with the Palatini $f(\mathfrak{R})$-term has been proposed~\cite{hybrid1}. The model was called the hybrid metric-Palatini f(R)-gravity. This theory was investigated in scalar-tensor representation. It was shown that the hybrid f(R)-gravity allows to describe a cosmological large-scale  structure, without affecting the Solar System dynamics. Such results led to numerous studies of hybrid f(R)-gravity. The cosmological implication of this model was studied in many works. For example, static Einstein Universe~\cite{bohmer} and various cosmological models~\cite{lima, lea} were investigated, cosmological solutions were obtained and late-time cosmic acceleration was described~\cite{capo3}. Moreover, the hybrid f(R)-gravity was studied on astrophysical scales from stars to galaxy clusters. It was shown, that the virial mass discrepancy in clusters of galaxies can be explained via the geometric terms appearing in the generalized virial theorem~\cite{capo4}. The hybrid f(R)-gravity also allows to explain the rotational velocities of test particles gravitating around galaxies. This approach allows to avoid introducing of a huge amount of dark matter~\cite{capo5}. Besides, wormhole solutions~\cite{capo6} were derived and physical properties of neutron, Bose-Einstein Condensate and quark stars were considered~\cite{stars}. Thus, hybrid f(R)-gravity appears to be very perspective.
	
	The main conceptual reason for introducing the hybrid f(R)-gravity is the following. As discussed in detail earlier, if f(R)-gravity is represented in a scalar-tensor form, then the metric f(R)-model corresponds to the Brans-Dicke theory with the parameter $\omega_{BD}=0$, whereas Palatini f(R)-gravity corresponds to the model with $\omega_{BD}=-3/2$. Both options are incompatible with the restrictions imposed by Solar System observations, since the original Brans-Dicke theory predicts $\omega_{BD}\to\infty$. This discrepancy is overcome if we consider such an action, in which the standard part of GR, i.e., $R$, is determined according to the metric approach, while the further degrees of freedom of the gravitational field, i.e., $f(\mathfrak{R})$-term defined by the Palatini method. In this case, the scalar field is dynamical, and the shortcomings of both the metric and the Palatini models are overcome. An attractive feature of the theory is that it allows non-standard scalar-tensor representation in terms of a dynamical scalar field (unlike the Palatini models), which does not have to be very massive to be consistent with the data obtained from laboratory experiments and Solar System observations. The features in the evolution of cosmological perturbations that appear in the Palatini models do not arise in this theory, because the scalar field is very weakly coupled to matter. Therefore, in this theory, the scalar field can play an active role in cosmology, without entering into the conflict with local experiments. We refer the reader to the work~\cite {hybrid2} for a review of the motivations for introducing hybrid f(R)-gravity.
	
	Previously, the effects of hybrid f(R)-gravity were studied both on cosmological scales and in the weak field limit. The theory was constrained based on experiments on the determination of the gravitational constant and the deflection of light in the gravitational field of the Sun (Cassini experiment)~\cite{lea}. However, the most complete test of any gravitational theory  in Solar System is the parameterized post-Newtonian formalism. Our work is dedicated to this issue. Main goal is to constrain the theory, taking into account the latest most accurate values of the PPN parameters ($\gamma$ and $\beta$) obtained from MESSENGER~\cite{mercury1, mercury2, mercury3, mercury4}.
	
	Parameterized post-Newtonian (PPN) formalism was developed  by  C. M. Will and K. Nordtvedt~\cite{PPN1,PPN2, ppn1, will1}. Within the PPN formalism metrics of various gravitational theories can be expressed as general PPN one which consists of PPN potentials and PPN parameters. Distinctions between gravitational models and observations are reflected via the set of 10 PPN parameters while the PPN potentials should be the same. 
	
	The picture changes when the theory implies the presence of massive fields. In this case the metric includes not only standard PPN potentials but also the Yukawa-type ones. Thus, the PPN formalism cannot be directly applied to gravitational models with massive fields~\cite{massbd}. There are two ways to modify the PPN formalism to make it suitable for testing such theories. The first one to introduce new PPN potentials including Yukawa-type ones. In this case, the new PPN parameters remain constants, however, it is necessary to provide their connection with the standard PPN ones and with the Solar System experiments~\cite{helbig}. The second possibility is the preservation of the standard form of PPN potentials with the introduction of spatial dependence in PPN parameters~\cite{massivebd}. Then PPN parameters cease to be constants, lose universality and their experimental values depend on the distance at which they are measured. Nevertheless, this method is also acceptable for testing and constraining gravitational theories with massive fields at a specific distance. In this paper we use the second way of PPN formalism modification and impose restrictions of hybrid f(R)-gravity in the weak-field limit. 
	
	The structure of the paper is the following. In section~\ref{sec2} we consider the action and the field equations of the hybrid metric-Palatini theory in a general form and in a scalar-tensor representation.  In section~\ref{sec4}, we discuss PPN formalism, solve the post-Newtonian equations of the hybrid f(R)-gravity and obtain the analytical expressions for the effective PPN parameters. Further, in section~\ref{sec5} we impose restrictions on the hybrid f(R)-gravity using the observational values for PPN parameters. We conclude in section~\ref{sec:conclusions} with a summary and discussion.
	
	Throughout this paper the Greek indices $(\mu, \nu,...)$ run over $0, 1, 2, 3$ and the signature is  $(-,+,+,+)$. All calculations are performed in the CGS system.
	
	\section{Hybrid f(R)-gravity}\label{sec2}
	
	In this section we discuss the main features of the hybrid f(R)-theory. The action is specified as~\cite{hybrid1,hybrid2}
	\begin{equation}\label{act}
	S=\frac{c^4}{2k^2}\int d^4x\sqrt{-g}\left[R+f(\mathfrak{R})\right]+S_m,
	 \end{equation}
	where $c$ is the speed of light, $k^2=8\pi G$, $R$ and $\mathfrak{R}=g^{\mu\nu}\mathfrak{R}_{\mu\nu}$ are the metric and Palatini curvatures respectively, $g$ is the metric determinant, $S_m$ is the matter action. Here the Palatini curvature $\mathfrak{R}$ is defined as a function of $g_{\mu\nu}$ and the independent connection $\hat\Gamma^\alpha_{\mu\nu}$:
\begin{equation}\label{re}
	\mathfrak{R}=g^{\mu\nu}\mathfrak{R}_{\mu\nu}=g^{\mu\nu} \bigl(\hat\Gamma^\alpha_{\mu\nu,\alpha}-\hat\Gamma^\alpha_{\mu\alpha,\nu}+\hat\Gamma^\alpha_{\alpha\lambda}\hat\Gamma^\lambda_{\mu\nu}-\hat\Gamma^\alpha_{\mu\lambda}\hat\Gamma^\lambda_{\alpha\nu} \bigr).
	 \end{equation}
	
	Like in the pure metric and Palatini cases, the hybrid f(R)-gravity~(\ref{act}) can be rewritten in a scalar-tensor representation (for details see~\cite{hybrid1,hybrid2})
	\begin{equation}\label{stact1}
	S=\frac{c^4}{2k^2}\int d^4x\sqrt{-g}  \biggl[(1 + \phi)R + \frac{3}{2\phi}\partial_\mu \phi \partial^\mu \phi - V(\phi)  \biggr]+S_m,
	 \end{equation}
	where $\phi$ is a scalar field and $V(\phi)$ is a scalar field potential. Here and further we use the Jordan frame.
	
	After variation and several transformations it is possible to obtain metric and scalar field equations respectively~\cite{hybrid1,hybrid2}
	\begin{wide}
	\begin{eqnarray}
 	&&(1+\phi)R_{\mu\nu}=\frac{k^2}{c^4}\left(T_{\mu\nu}-\frac{1}{2}g_{\mu\nu}T\right)-\frac{3}{2\phi}\partial_\mu\phi\partial_\nu\phi +\frac{1}{2}g_{\mu\nu}  \biggl[V(\phi)+\nabla_\alpha\nabla^\alpha\phi  \biggr]+\nabla_\mu\nabla_\nu\phi,\label{feh}\\
 	&&\nabla_\mu\nabla^\mu\phi-\frac{1}{2\phi}\partial_\mu\phi\partial^\mu\phi-\frac{\phi[2V(\phi)-(1+\phi)V_\phi]}{3}=-\frac{k^2}{3c^4}\phi T.\label{fephi}
	 \end{eqnarray}
	 \end{wide}
	Unlike the Palatini case, in the hybrid f(R)-gravity the scalar field is dynamical. Thus, the theory is not affected by the microscopic instabilities arising in Palatini models~\cite{hybrid1,hybrid2}.
	
	\section{PPN-limit of hybrid f(R)-theory}\label{sec4}
	
	PPN formalism was created to compare different gravitational theories with each other and experiments~\cite{will1}. The post-Newtonian (PN) limit is achieved in approximation of slow velocities, asymptotically flat space-time background and weak gravitational field. Thus, it allows to test gravitational models in the Solar System with a high accuracy.
	
	Originally C. M. Will and K. Nordtvedt developed slightly different approaches~\cite{PPN1,PPN2}. K. Nordtvedt investigated the post-Newtonian metric for point-mass gravitational systems, while Will considered the matter in the perfect fluid approximation~\cite{PPN1,PPN2}. Later it was shown that both methods are equivalent~\cite{ppn1}. In this paper the Nordtvedt approach is applied. 
	
	To consider the hybrid f(R)-theory in the weak-field limit we expand the scalar $\phi$ and tensor $g_{\mu\nu}$ fields as
	\begin{equation}\label{decompos}
	\phi=\phi_0+\varphi,\qquad\ g_{\mu\nu}= \eta_{\mu\nu}+h_{\mu\nu},
	 \end{equation}
	where $\phi_0$ is the asymptotic background value of the scalar field far away from the source, $\eta_{\mu\nu}$ is the Minkowski background, $h_{\mu\nu}$ and $\varphi$ are the small perturbations of tensor and scalar fields respectively. In the general case $\phi_0$ is not a constant but the function of time $\phi(t)$. However this dependence can be neglected whenever its characteristic time scale is very long compared with the dynamical time scale associated with the local system itself. Thus, $\phi_0$ is taken as a constant.
	
	The complete post-Newtonian limit requires to evaluate the different components of the metric and the scalar field perturbations to the following orders $h_{00}\sim O(2) + O(4), h_{0j}\sim O(3), h_{ij}\sim O(2)$ and $\varphi\sim O(2) + O(4)$ (see~\cite{will1}). Thus, the scalar potential $V(\phi)$ could be expanded in a Taylor series around the background value of scalar field $\phi_0$ like
	\begin{equation}\label{V}
	V(\phi)=V_0+V'\varphi+\frac{V''\varphi^2}{2!}+\frac{V'''\varphi^3}{3!}...
	 \end{equation}
	hence its derivative with respect to $\varphi$ will take the form $V_\phi=V'+V''\varphi+V'''\varphi^2/2$.
	
	The energy-momentum tensor for point-mass gravitational system is defined as
	\begin{equation}\label{emt1}
	T^{\mu\nu}=\frac{c}{\sqrt{-g}}\sum_a m_a\frac{u^{\mu}u^{\nu}}{u^0} \delta^3( \mathbf{r}- \mathbf{r}_a),
	 \end{equation}
	where $m_a$ is the mass of the $a$-th particle, $\mathbf{r}_a$ is the radius-vector of the $a$-th particle, $u^{\mu}=d x^{\mu}_a/d \tau_a$ is four-velocity of the $a$-th particle, $d\tau=\sqrt{-ds^2}/c$, $ds^2=g_{\mu\nu}dx^{\mu}dx^{\nu}$ is an interval, $u_{\mu}u^{\mu}=-c^2$, and $\delta^3(\mathbf{r}-\mathbf{r}_a(t))$ is the three-dimensional Dirac delta function.
	
	In the post-Newtonian approximation components of the energy-momentum tensor~(\ref{emt1}) and its trace take the form:
	\begin{eqnarray}\label{emt}
		T_{00}&=&c^2\sum_a m_a \delta^3( \mathbf{r}- \mathbf{r_a})\left[1-\frac{3}{2}h_{00}+\frac{1}{2}\frac{v^2_a}{c^2}-\frac{1}{2}h\right],\\
	T_{0i}&=&-c\sum_am_av_a^i \delta^3( \mathbf{r}- \mathbf{r_a}),\\
	T_{ij}&=&\sum_am_av_a^iv_a^j \delta^3( \mathbf{r}- \mathbf{r_a}),\\
	T&=&-c^2\sum_am_a \delta^3( \mathbf{r}- \mathbf{r_a})\left[1-\frac{1}{2}h_{00}-\frac{1}{2}\frac{v_a^2}{c^2}-\frac{1}{2}h\right],
	 \end{eqnarray}
	where $v_a$ is the velocity of the $a$-th particle.
	
	To obtain the field equations~(\ref{feh}) and~(\ref{fephi}) in the weak-field limit~(\ref{decompos}) we apply the Nutku gauge conditions~\cite{Nutku}:
	\begin{equation}\label{gauge}
	h^\alpha_{\beta,\alpha}-\frac{1}{2} \delta^\alpha_\beta h^\mu_{\mu,\alpha}=\frac{\varphi_{,\beta}}{1+\phi_0}.
	 \end{equation}
	
	\subsection{Solutions for $\varphi^{(2)}$, $h_{00}^{(2)}$, $h_{ij}^{(2)}$}\label{sec4_1}
	
	The consideration starts with the obtaining of Newtonian limit for  the hybrid f(R)-gravity. The field equation for the scalar field~(\ref{fephi}) in the leading perturbation order ($O(2)$) takes the form
	\begin{equation}\label{fephi2}
	\left(\nabla^2-m_\varphi^2\right)\varphi^{(2)}=\frac{k^2\phi_0}{3c^2}\sum_am_a \delta^3( \mathbf{r}- \mathbf{r_a}),
	 \end{equation}
	where we denote $m_\varphi^2=[2V_0-V'-(1+\phi_0)\phi_0V'']/3$ as a scalar field mass. The zeroth-order term $\phi_0[2V_0-(1+\phi_0)V']/3$ which should appear in the scalar field equation can be absorbed into a coordinate redefinition. The superscript $^{(2)}$ indicates the order of perturbation.
	
	Using the general solution of the screened Poisson equation and properties of the Dirac delta function we obtain
	\begin{equation}\label{phi22}
	\varphi^{(2)}=-\frac{k^2\phi_0}{12\pi c^2}\sum_am_a\frac{ e^{-m_\varphi r_a}}{r_a},
	 \end{equation}
	where $r_a=|\mathbf{r}-\mathbf{r}_a|$.
	
	The 00-component of~(\ref{feh}) up to the order $O(2)$ is given by
	\begin{eqnarray}\label{feh002}
	\nabla^2\left(h_{00}^{(2)}-\frac{\varphi^{(2)}}{1+\phi_0}\right)=&-&\frac{k^2}{c^2(1+\phi_0)}\sum_am_a \delta^3( \mathbf{r}- \mathbf{r_a})\nonumber\\
	&+&\frac{V_0}{1+\phi_0}.
	 \end{eqnarray}
	Using derived expression for $\varphi^{(2)}$~(\ref{phi22}) and assuming that the main contribution to the metric in the Solar System is due to the Sun the solution for $h_{00}^{(2)}$ can be represented as
	\begin{equation}\label{h002_1}
	h_{00}^{(2)}=\frac{k^2}{4\pi(1+\phi_0)c^2}\frac{M}{r}\left(1-\frac{\phi_0}{3}   e^{-m_\varphi r}\right)+\frac{V_0}{1+\phi_0}\frac{r^2}{6},
	 \end{equation}
where $M$ is the Solar mass. Here $V_0/(\phi_0+1)$ is the cosmological constant term. It must be negligible in Solar System scales in order not to affect the local dynamics. That is why further this term is not considered.
	
From~(\ref{h002_1}) it is possible to extract the effective gravitational constant~\cite{hybrid1,hybrid2}:
	\begin{equation}\label{Geff}
	G^{eff}=\frac{k^2}{8\pi(1+\phi_0)}\left(1-\frac{\phi_0}{3}   e^{-m_\varphi r}\right).
	 \end{equation}
Throughout the article we add the superscript $^{\rm eff}$ to the PPN parameters which are considered as spatially dependent functions. Another superscript $^{\rm exp}$ is used for the experimental values of PPN parameters. When we mean the PPN parameters entered in the original PPN formalism, we do not use indices. These superscripts are also applied to the gravitational constant.
		
In the hybrid f(R)-gravity the effective gravitational constant is not a constant actually, but it is the function of a distance. The Newtonian limit can be reproduced in two ways: $\phi_0\ll1$ or $m_\varphi r\gg 1$. The first case implies the possibility of the existence of a light long-range scalar field. Therefore, it is not necessary to use a screening mechanism for description of the Solar System dynamics in the hybrid f(R)-gravity~\cite{hybrid1,hybrid2}. 

The tensor field equation for the $ij$-component is
	\begin{eqnarray}\label{fehij}
	\nabla^2\left(h_{ij}^{(2)}+ \delta_{ij}\frac{\varphi^{(2)}}{1+\phi_0}\right)&=&-\biggl(\frac{k^2}{(1+\phi_0)c^2} \sum_am_a \delta^3( \mathbf{r}- \mathbf{r_a})\nonumber\\
	&&- \frac{V_0}{1+\phi_0}\biggr)\delta_{ij},
	 \end{eqnarray}
where $\delta_{ij}$ is the Kronecker delta. We obtain the solution analogically to $h_{00}^{(2)}$
		\begin{equation}\label{hij_1}
	h_{ij}^{(2)}=\frac{ \delta_{ij}k^2}{4\pi(1+\phi_0)c^2}\frac{M}{r}\left(1+\frac{\phi_0}{3}   e^{-m_\varphi r}\right)- \delta_{ij}\frac{V_0}{1+\phi_0}\frac{r^2}{6}.
	 \end{equation}
	
To extract the modified PPN parameters, the obtained metric should be compared with the general point-mass one~(\ref{pmmet}) introduced by K. Nordtvedt~\cite{ppn1} (see~\ref{app1}).
	
After comparing~(\ref{hij_1}) with~(\ref{pmmet}), the effective PPN parameter $\gamma^{\rm eff}$ can be expressed as~\cite{hybrid1,hybrid2}
	\begin{equation}\label{gamma}
	\gamma^{eff}=\frac{1+\phi_0   e^{-m_\varphi r}/3}{1-\phi_0   e^{-m_\varphi r}/3}.
	 \end{equation}
Note that $\gamma^{\rm eff}$ is spatially dependent. The current observations predict that $\gamma^{\rm exp}\approx 1$ (in GR $\gamma=1$) with high accuracy~\cite{gamma,mercury1,mercury2,mercury3,mercury4, will14}. One way to achieve this result is the consideration of the case  $\phi_0\ll1$. Thus, the resulting expression for $\gamma^{\rm eff}$ does not contradict to the assumption that a scalar field can be light~\cite{hybrid1,hybrid2}. 
	
Besides, there is another way to obtain the expression for $\gamma$ from the solution of the light propagation equation. In~\cite{massbd} it was obtained in detail for massive Brans-Dicke theory. The crucial point was an inequality of the observed Keplerian mass and mass of the body causing the time-delay. Since hybrid metric-Palatini f(R)-gravity can be represented as massive scalar-tensor model, an expression for $\gamma$ can be obtained in the same way. The result received by such method is identical to~(\ref{gamma}), which means there is no difference how to calculate this PPN parameter.
	
\subsection{Solutions for $\varphi^{(4)}$, $h_{00}^{(4)}$}\label{sec4_2}
	
After applying the expansions~(\ref{decompos}) up to $O(4)$ order, the field equations~(\ref{feh}) and~(\ref{fephi}) take the form
	\begin{wide}
	\begin{eqnarray}\label{weakfield}
	 	(\nabla^2-m^2_\varphi)\varphi^{(4)}=&&\frac{k^2\phi_0}{3c^2}\sum_am_a \delta^3( \mathbf{r}- \mathbf{r_a}) \biggl[-\frac{1}{2}h^{(2)}_{jj}-\frac{1}{2}\frac{v^2}{c^2}  \biggr]+h_{ij}^{(2)}\varphi^{(2)}_{,ij}
	 	+\frac{k^2}{3c^2}\varphi^{(2)}\sum_am_a \delta^3( \mathbf{r}- \mathbf{r_a})+\varphi^{(2)}_{,00}\nonumber\\
	 	&+&\frac{3\phi_0+1}{2\phi_0(1+\phi_0)}(\nabla\varphi^{(2)})^2-\frac{(\varphi^{(2)})^2}{3} \biggl[\frac{V'''\phi_0(\phi_0+1)}{2}+V''(\phi_0+1)  \biggr],
	 \end{eqnarray}
	\begin{eqnarray}\label{weakfield00}
 	\nabla^2  \biggl(h^{(4)}_{00}-\frac{\varphi^{(4)}}{1+\phi_0}  \biggr)=&-&\frac{k^2}{c^2(1+\phi_0)}\sum_am_a \delta^3( \mathbf{r}- \mathbf{r_a})  \biggl[-h_{00}^{(2)}+\frac{3}{2}\frac{v^2_a}{c^2}-\frac{1}{2}h_{j}^j  \biggr]+h^{(2)}_{00,00}-(\nabla h^{(2)}_{00})^2+h^{(2)}_{ij}h^{(2)}_{00,ij}\nonumber\\
 	&-&\frac{1}{1+\phi_0}h^{(2)}_{ij}\varphi^{(2)}_{,ij}-\frac{\varphi^{(2)}_{,00}}{1+\phi_0}-\frac{1}{1+\phi_0}h_{00} \delta\varphi^{(2)}-\frac{\varphi^{(2)}}{1+\phi_0} \delta h^{(2)}_{00}-\frac{1}{(1+\phi_0)^2}(\nabla\varphi)^2.
	 \end{eqnarray}
	\end{wide}
	
In order to provide a reasonable cosmological picture, $V_0$ should be of the same order as the energy density of the cosmological constant. Additionally we expect $V'(\phi_0)$ to be small enough so that its contribution can be considered negligible. The reason is based on the assumption that either the field approaches a minimum at late times, or the potential takes the form $V = V_0e^{-ak\phi}$ (where $a$ is of order unity), therefore $V'(\phi_0)\sim kV_0$. This assumption seems to be plausible in all reasonable models \cite{satirou}. Hence we neglect the terms containing $V_0$, $V'$ multiplied by perturbed quantities (e.g. $V_0h_{00}$), because these terms should not lead to any observable deviations at Solar System scales.
	
To solve~(\ref{weakfield}) and~(\ref{weakfield00}) it is convenient to use the following ratios:
	\begin{eqnarray}
	(\nabla\varphi)^2&=&\frac{1}{2}(\nabla^2-m_\varphi^2)\varphi^2-\varphi  \biggr(\nabla^2-\frac{m_\varphi^2}{2}  \biggl)\varphi,\\
	(\nabla h_{00})^2&=&\frac{1}{2}\nabla^2h_{00}^2-h_{00}\nabla^2h_{00}.
	 \end{eqnarray}
	
	Further, from~(\ref{weakfield}) for $\varphi^{(4)}$ we find:
	\begin{wide}
	\begin{eqnarray}\label{phi4}
 	\varphi^{(4)}=&&\frac{k^2\phi_0}{24\pi c^2}\sum_am_a\partial_t\partial_t\frac{ e^{-m_\varphi r_a}}{m_\varphi}
 	+\frac{k^4\phi_0(1+3\phi_0)}{576\pi^2c^4(1+\phi_0)}\sum_am_a\frac{ e^{-m_\varphi r_a}}{r_a}\sum_bm_b\frac{ e^{-m_\varphi r_b}}{r_b}\nonumber\\
 	&+&\frac{k^4\phi_0}{96\pi^2c^4(1+\phi_0)}\sum_a\sum_{b\neq a}\frac{m_am_b}{r_ar_{ab}} e^{-m_\varphi r_a}\left(1+\frac{\phi_0}{3} e^{-m_\varphi r_{ab}}\right)-\frac{k^4\phi_0(\phi_0-1)}{288\pi^2c^4(1+\phi_0)}\sum_a\sum_{b\neq a}\frac{m_am_b}{r_ar_{ab}} e^{-m_\varphi r_a} e^{-m_\varphi r_{ab}}\nonumber\\
&+&\frac{k^2\phi_0}{24\pi c^4}\sum_av_a^2\frac{m_a}{r_a} e^{-m_\varphi r_a}-\frac{k^4\phi_0m_\varphi}{96\pi^2c^4(1+\phi_0)}\sum_a\sum_{b\neq a}\frac{m_am_b}{r_{ab}}  \bigg[-Ei(-2m_\varphi r_a) e^{-m_\varphi r_{ab}} e^{m_\varphi r_a}+Ei(-2m_\varphi r_b) e^{m_\varphi r_b}\nonumber\\
&-&\ln(r_a) e^{-m_\varphi r_b}+\ln(r_b) e^{-m_\varphi r_b}  \bigg]-  \bigg[\frac{k^4\phi_0(1+7\phi_0)m_\varphi}{1152\pi^2c^4(1+\phi_0)}+\frac{k^4\phi_0^2(1+\phi_0)}{864\pi^2c^4m_\varphi} \bigg(V''+\phi_0\frac{V'''}{2}  \bigg)  \bigg] \sum_a\sum_{b\neq a}\frac{m_am_b}{r_{ab}}\nonumber\\
&\times&  \bigg[Ei(-3m_\varphi r_b) e^{4m_\varphi r_{ab}} e^{m_\varphi r_a}-Ei(-3m_\varphi r_a)  e^{m_\varphi r_{ab}} e^{m_\varphi r_a}-Ei(-m_\varphi r_b) e^{-m_\varphi r_a}+Ei(-m_\varphi r_a) e^{-m_\varphi r_b}  \bigg].
	 \end{eqnarray}
	 \end{wide}
	Here $Ei$ denotes the exponential integral defined as
	\begin{equation}\label{ei}
	Ei(-x)=-\int^{\infty}_x\frac{ e^{-t}}{t}dt.
	 \end{equation}
	
Thus, using the  solution~(\ref{phi4}) and expressions for $\varphi^{(2)}, h_{00}^{(2)}, h_{ij}^{(2)}$, from~(\ref{weakfield00}) we obtain 
\begin{wide}
\begin{eqnarray}\label{h004_1}
 	h_{00}^{(4)}=&-&\frac{k^4}{32\pi^2 c^4(1+\phi_0)^2}\sum_a\frac{m_a}{r_a}\left(1-\frac{\phi_0}{3} e^{-m_\varphi r_a}\right)\sum_b\frac{m_b}{r_b}\left(1-\frac{\phi_0}{3} e^{-m_\varphi r_b}\right)+\frac{k^4\phi_0(1+\phi_0)}{576\pi^2 c^4(1+\phi_0)^2}\nonumber\\
 	&\times&\sum_am_a\frac{ e^{-m_\varphi r_a}}{r_a}\sum_bm_b\frac{ e^{-m_\varphi r_b}}{r_b}-\frac{k^4}{32\pi^2 c^4(1+\phi_0)^2}\sum_a\sum_{b\neq a}\frac{m_am_b}{r_ar_{ab}}  \biggl(1-\frac{\phi_0}{3} e^{-m_\varphi r_a}  \biggr)\left(1-\frac{\phi_0}{3} e^{-m_\varphi r_{ab}}\right)\nonumber\\
 	&+&\frac{k^4\phi_0(\phi_0+1)}{288\pi^2 c^4(1+\phi_0)^2}\sum_a\sum_{b\neq a}\frac{m_am_b}{r_ar_{ab}} e^{-m_\varphi r_a} e^{-m_\varphi r_{ab}}+\frac{k^2}{8\pi c^4(1+\phi_0)}\sum_av_a^2\frac{m_a}{r_a}  \biggl(1-\frac{\phi_0}{3} e^{-m_\varphi r_a}  \biggr)\nonumber\\
&+&\frac{k^2}{4\pi c^4(1+\phi_0)}\sum_av_a^2\frac{m_a}{r_a}  \biggl(1+\frac{\phi_0}{3} e^{-m_\varphi r_a}  \biggr)-\frac{k^4\phi_0m_\varphi}{96\pi^2 c^4(1+\phi_0)^2}\sum_a\sum_{b\neq a}\frac{m_am_b}{r_ar_{ab}}  \bigg[-Ei(-2m_\varphi r_a) r_a e^{-m_\varphi r_{ab}} e^{m_\varphi r_a}\nonumber\\
&+&Ei(-2m_\varphi r_b)r_a e^{m_\varphi r_b}-\ln(r_a)r_a e^{-m_\varphi r_b}	+\ln(r_b)r_a e^{-m_\varphi r_b}  \bigg]-  \bigg[\frac{k^4\phi_0(1+7\phi_0)m_\varphi}{1152\pi^2c^4(1+\phi_0)^2}\nonumber\\
&+&\frac{k^4\phi_0^2}{864\pi^2c^4m_\varphi}  \bigg(V''+\phi_0\frac{V'''}{2}  \bigg)  \bigg]\sum_a\sum_{b\neq a}\frac{m_am_b}{r_ar_{ab}}  \bigg[Ei(-3m_\varphi r_b)r_a e^{4m_\varphi r_{ab}} e^{m_\varphi r_a}-Ei(-m_\varphi r_b)r_a e^{-m_\varphi r_a}\nonumber\\
&+&Ei(-m_\varphi r_a)r_a e^{-m_\varphi r_b}-Ei(-3m_\varphi r_a)r_a e^{m_\varphi r_{ab}} e^{m_\varphi r_a}  \bigg]+\frac{k^2}{4\pi c^2(1+\phi_0)}\sum_am_a\partial_t\partial_t\left(\frac{r_a}{2}\right)\nonumber\\
&+&\frac{k^2\phi_0}{24\pi c^2(1+\phi_0)}\sum_am_a\partial_t\partial_t\left(\frac{ e^{-m_\varphi r_a}}{m_\varphi}\right).
 \end{eqnarray}
\end{wide}
Comparing the PPN-metric for hybrid f(R)-gravity with the general Nordtvedt point-mass one~(\ref{pmmet}), it is possible to extract effective PPN parameters. The analytical expression for $\beta^{\rm eff}$ can be obtained from the first two terms~(\ref{h004_1}). After adding the natural assumption that the Sun provides the main contribution in the Solar System $\beta^{\rm eff}$ takes the form:
	\begin{equation}\label{beta}
	\beta^{eff}=1-\frac{\phi_0(\phi_0+1) e^{-2m_\varphi r}}{18(1-\frac{\phi_0}{3} e^{-m_\varphi r})^2}.
	 \end{equation}
Thus the parameter $\beta^{\rm eff}$ (as well as $\gamma^{\rm eff}$) assumes two variants at which value $\beta^{\rm eff}\approx 1$ (GR value $\beta=1$) is reached: $\phi_0\ll1$ or $m_\varphi r\gg 1$. 

Considering the terms like $\sum_av_a^2\frac{m_a}{r_a}$ it is possible to identify the effective PPN parameters $\alpha_3^{\rm eff}=\zeta_1^{\rm eff}=0$. The terms like $\sum_a\sum_{b\neq a}\frac{m_am_b}{r_ar_{ab}}$ are accompanied by the combination $-2\beta^{\rm eff}+1+\zeta_2^{\rm eff}$ (see~(\ref{pmmet})). After extracting terms with already known effective PPN parameters only the contributions multiplied by $m_\varphi,  V'', V'''$ remain. All these terms should contribute to $\zeta^{\rm eff}_2$. Here it is important to emphasize that in hybrid f(R)-gravity $\alpha_3=\zeta_1=\zeta_2=\zeta_3=\zeta_4$ are equal to zero~\cite{lee} because the usual conservation laws are satisfied~\cite{hybrid1,hybrid2}. Thus, all contributions multiplied by $m_\varphi,  V'', V'''$ should appear in the next PPN order so it is possible to neglect them.
	
In~(\ref{h004_1}) only terms with temporal derivatives remain. We discuss these contributions in the next subsection.
	
\subsection{Solutions for $h_{0i}^{(3)}$}
In this subsection the field equations~(\ref{feh}) up to the order $O(3)$ are discussed. The solution is 
	\begin{equation}\label{h0i_1}
	h_{0i}^{(3)}=-\frac{k^2}{2(1+\phi_0)c^3} \sum_a\frac{m_a}{r_a}v^i_a.
	 \end{equation}
	
In this paper we use the conformal harmonic gauge~(\ref{gauge}). To bring it to the standard post-Newtonian gauge we implement the coordinate transformation $t = \bar t + \partial_{\bar t}X/2c^4$ and $x^j=\bar x^j$. Here $X$ is the superpotential defined as $\nabla^2 X=2G^{\rm eff} M/r$. Transforming the metric to the new coordinates $(\bar t, \bar x^j )$ and droping the overbars on the new variables, we obtain that the solution for $ij$-component would be the same. In the $00$-component all terms with temporal derivatives vanish, while the $0i$-component takes the form~\cite{Will}
	\begin{wide}
	\begin{eqnarray}\label{h0i_2}
	h_{0i}^{(3)}=&-&\frac{3k^2}{16\pi c^3(1+\phi_0)}\sum_{a}\frac{m_av_a^i}{r_a}  \biggl(1-\frac{\phi_0}{3} e^{-m_\varphi r_a}  \biggr)-\frac{k^2}{4\pi c^3(1+\phi_0)} \sum_{a}\frac{m_av_a^i}{r_a}  \biggl(1+\frac{\phi_0}{3} e^{-m_\varphi r_a}  \biggr)\nonumber\\
&+&\frac{k^2}{16 \pi c^3(1+\phi_0)}\sum_{a}\frac{m_ar_a^i}{r_a^3}( \mathbf{v}_a   \mathbf{r}_a)  \biggl(1-\frac{\phi_0}{3} e^{-m_\varphi r_a}  \biggr).
	 \end{eqnarray}
	\end{wide}
	
After comparing this expression with the general point-mass metric~(\ref{pmmet}) it is possible to conclude that $\alpha^{\rm eff}_1=\alpha^{\rm eff}_2=0$. Thus, there are no preferred-frame effects in the hybrid f(R)-gravity.
	
\subsection{Perfect fluid metric}

Besides the point-mass metric, we obtain also the perfect fluid PPN metric of the hybrid f(R)-gravity~(\ref{metr}). We place it in the~\ref{app2} because of its length. Here we present only a discussion. 
	
In the original PPN formalism all PPN parameters are constants because it was developed for massless theories. However, an application of the PPN formalism to gravitational theories with massive fields leads to appearance of spatially dependent PPN parameters. Considering the hybrid f(R)-gravity in the perfect fluid approximation we find that PPN parameters are not only spatial dependent functions but also a part of PPN potentials. Therefore their identification becomes difficult and their physical meaning is unclear. However some details can be extracted from this metric.
	
Original perfect fluid metric~(\ref{pfmet}) includes a set of ten PPN parameters: $\gamma, \beta, \xi, \zeta_{1,2,3,4}, \alpha_{1,2,3}$. They are equivalent to PPN parameters of point-mass metric~(\ref{pmmet}) but  $\xi$ and $\zeta_{3,4}$ are not included in the last one. However, comparing the obtained metric~(\ref{metr}) with the general perfect fluid one~(\ref{pfmet}) it is possible to find that $\xi^{\rm eff}=0,\zeta^{\rm eff}_3=0$ in the hybrid f(R)-gravity. The parameter $\zeta^{\rm eff}_4$ is the combination of other PPN parameters $6\zeta_4=3\alpha_3+2\zeta_1-3\zeta_3$~\cite{will14}. Since all parameters in the combination are equal to zero hence $\zeta^{\rm eff}_4=0$. 
	
\section{Observational limits}\label{sec5}

The main goal of this work is to impose restrictions on the hybrid f(R)-gravity and to study its behavior in the Solar System. To find the limitations on $\phi_0$ and $m_\varphi$ the MESSENGER data for $\gamma$ and $\beta$ is used~\cite{mercury3}. Recently, newer experimental data for $\beta$ were obtained \cite{newmes}. In this work authors combine $\gamma^{\rm exp}$ obtained during Cassini mission with measurements of secular and periodic precession of Mercury's orbit that allows to estimate both $\beta^{\rm exp}$ and $J_2$. Thus, the value of the $\beta^{\rm exp}$ was obtained via $\gamma^{\rm exp}$ which was found at a distance from the gravitating source (the Sun) different from the distance where the MESSENGER experiment was conducted. Therefore these new data for $\beta$  cannot be used to test massive scalar-tensor theories, since the PPN parameters are functions of $r$ and can vary depending on the distance at which they were measured.
	
\begin{figure}[t]
		\begin{minipage}[h]{.49\textwidth}
			\center{\includegraphics[width=1\linewidth]{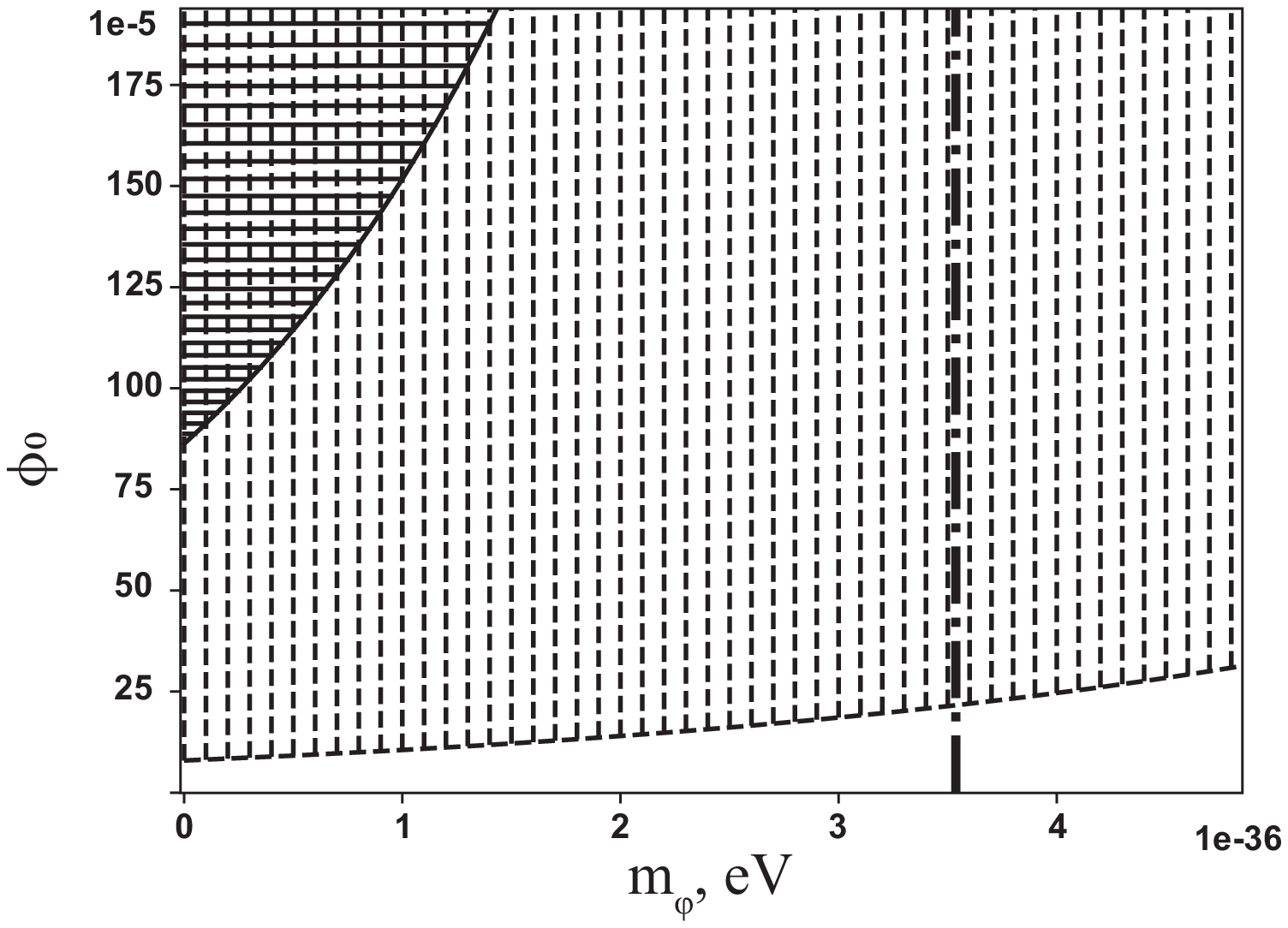}\\ a)}
			\end{minipage}
			\hfill
			\begin{minipage}[h]{.49\textwidth}
			\center{\includegraphics[width=1\linewidth]{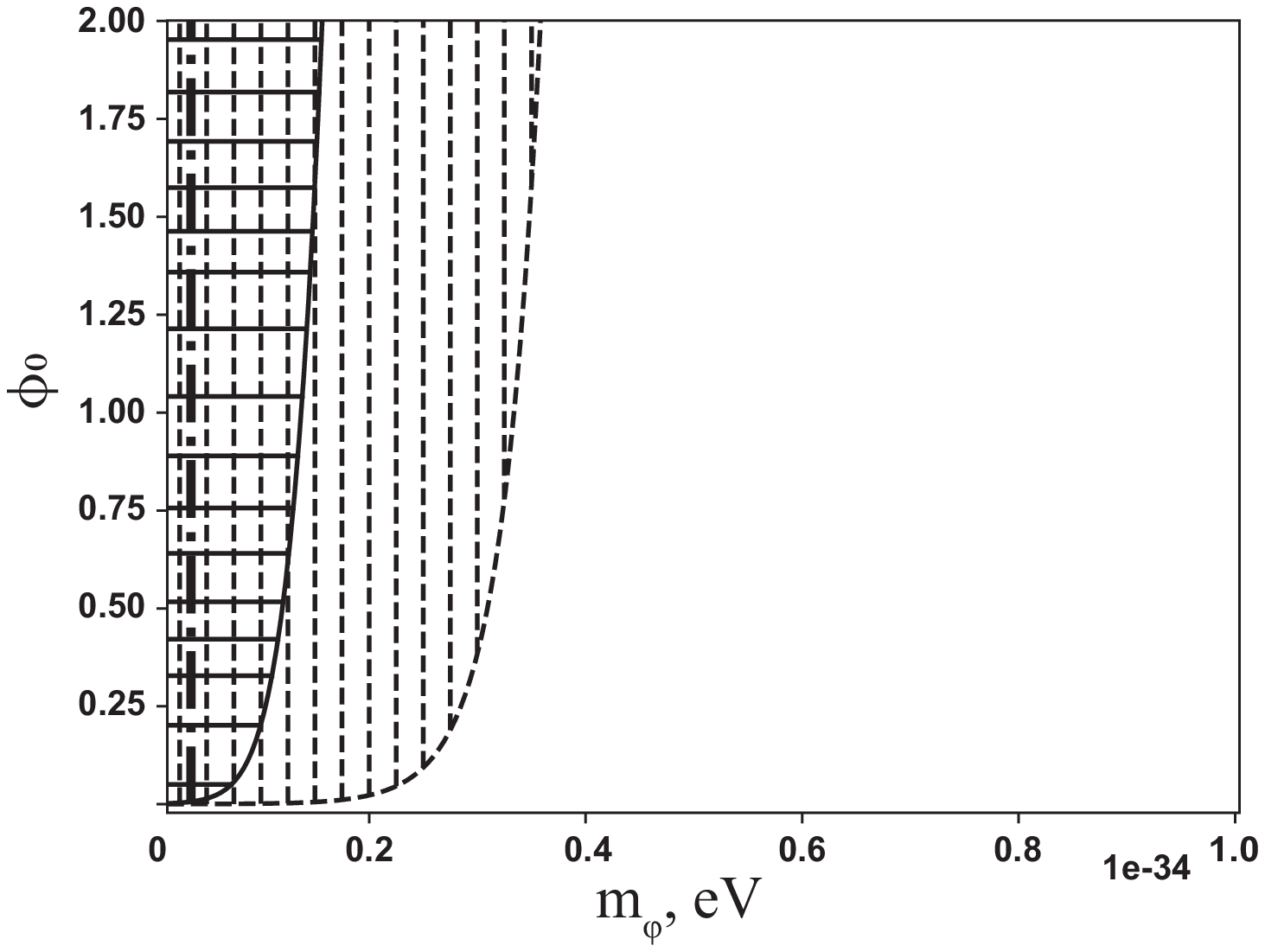} \\ b)}
			\end{minipage}
\caption{Dependence of scalar field background value upon the scalar field mass. Two figures show allowable regions at different scales. The vertical dotted region corresponds to excluded values obtained by  $\gamma^{\rm exp}$, the horizontal solid region corresponds to excluded values obtained by $\beta^{\rm exp}$, the vertical dash dotted  line is the critical value of scalar mass $m_\varphi=\frac{1}{r_0}$, where $r_0$ is equal to the distance from the Sun to Mercury.}
\label{hybridf3}
\end{figure}

In this work we constrain hybrid f(R)-gravity using the following experimental values $\gamma^{\rm exp}$ and $\beta^{\rm exp}$~\cite{mercury3,mercury1}: $\gamma^{\rm exp}=1-0.3\times10^{-5}\pm2.5\times10^{-5}$ and $\beta^{\rm exp}=1+0.2\times10^{-5}\pm2.5\times10^{-5}$. Restrictions on $\phi_0$ and $m_\varphi$ obtained from this data are shown in the figures~\ref{hybridf3}. Shaded areas reflect excluded regions. It is obvious that the $\gamma^{\rm exp}$ gives the best limits, compared with the $\beta^{\rm exp}$. It is demonstrated that for small values of $\phi_0$, the scalar mass can take any values, including very small ones. For large values of scalar mass $\phi_0$ takes any values. Next, we consider two limiting cases, which allow to constrain the value of $\phi_0$.

Firstly, consider the case of very light scalar field $m_\varphi r\ll1$. Then it is possible to restrict $\phi_0$ as
	\begin{equation}
	-8\times10^{-5}<\phi_0<7\times10^{-5}
	\end{equation}
from the $\gamma^{\rm exp}$ and
	\begin{equation}
	-9\times10^{-4}<\phi_0<9\times10^{-4}
	\end{equation}
from the $\beta^{\rm exp}$ at the $2\sigma$ confidence level. The constraints obtained from $\gamma^{\rm exp}$ are more stringent than those ones obtained from $\beta^{\rm exp}$. 
	
In the case of massive scalar field $\gamma^{\rm eff}\approx 1$ and $\beta^{\rm eff}\approx1$. Then $\phi_0$ can be constrained from $G^{\rm eff}$ and its experimental value~\cite{geff}. In~\cite{lea} it was shown that in such case $\phi_0$ has also a small value ($|\phi_0|<5\times10^{-4}$).
	
Thus, experimental data from the Solar System indicates that $\phi_0$ is close to zero. In this case, $m_\varphi$ can take any values, therefore, it is not possible to set constraints on the mass of the scalar field in the weak field limit at the current moment. However, restrictions on $m_\varphi$ were obtained from other local systems such as binary pulsars (for details, see \cite{we}).
	
\section{Conclusion}\label{sec:conclusions}
	
In this paper we consider the post-Newtonian limit of hybrid metric-Palatini f(R)-gravity. Since this theory can be represented as massive scalar-tensor model~\cite{hybrid1,hybrid2} the original parameterized post-Newtonian formalism~\cite{PPN1,PPN2,ppn1,will1,Will} is not directly applicable~\cite{massbd}. However, there are still two ways to use the PPN formalism for testing a gravitational theory with massive fields. The first one is to develop a modified PPN formalism including not only original PPN potentials but in addition Yukawa-type ones. In such approach the modified PPN parameters remain constants, but require redefinition to be coupled with the experiments~\cite{helbig}. The second way is to preserve the PPN potentials in their original form and to include all modifications related to the presence of massive fields in PPN parameters. In the last approach modified PPN parameters are not constants but spatially dependent functions~\cite{massivebd}. We consider the second way.
	
We present modified PPN metrics of hybrid f(R)-gravity in two different approximations of local system matter: point-mass and perfect fluid~\cite{ppn1}. Using the first approach we extract the effective PPN parameters, since in the second one they are not only spatial dependent functions but also a part of PPN potentials. Therefore their extraction is difficult and moreover their physical meaning is uncertain. Thus, to use perfect fluid approximation it is necessary to modify original PPN potentials leaving the PPN parameters as constants (redefining them) to test gravitational theories with massive fields in weak-field limit.
	
We obtain the expressions for 10 effective PPN parameters in hybrid f(R)-gravity. Only $\gamma^{\rm eff}$ and $\beta^{\rm eff}$ are not trivial. Parameter $\gamma^{\rm eff}$ has been investigated earlier~\cite{hybrid1,hybrid2} since the expression for $\beta^{\rm eff}$ is received for the first time. As well as $\gamma^{\rm eff}$, the expected value $\beta^{\rm eff}\approx 1$ reaches in two cases: $\phi_0\ll1$ or $m_\varphi r\gg 1$. The first one allows the scalar field to be very light leaving Solar System unaffected but modifying cosmological and galactic dynamics without introducing screening mechanisms. Moreover, it was previously shown that even in the case of a very massive scalar field, the background value $\phi_0$ should still remain small ($|\phi_0|<5\times10^{-4}$)~\cite{lea}. This test is based on the effective gravitational constant of hybrid f(R)-gravity. In the assumption that scalar field is light we impose restrictions on the background value $\phi_0$ from $\gamma^{\rm eff}$ and $\beta^{\rm eff}$ using the  MESSENGER data~\cite{mercury1,mercury2,mercury3,mercury4} at the $2\sigma$ confidence level. We found that constraints obtained from $\gamma^{\rm exp}$ are more stringent ($-8\times10^{-5}<\phi_0<7\times10^{-5}$). 
	
	It was previously shown that the light scalar field in hybrid f(R)-gravity does not contradict the observational data obtained in the Solar System. The conclusion was made on the basis of the only PPN parameter $\gamma$~\cite {hybrid2, lea}. In our work, we performed a complete post-Newtonian analysis and clearly showed that the light scalar field in hybrid f(R)-gravity does not contradict the experimental data based not only on the $\gamma$ parameter, but also on all other parameters of the post-Newtonian formalism.

Despite the fact that hybrid f(R)-gravity does not contradict the observations in the weak-field limit it will be interesting to test the theory in the strong field regime of binary pulsars. Some constraints  have already been obtained  by testing the hybrid f(R)-gravity on the observational data of the orbital period changes in the systems PSR J1738+0333, PSR J0737-3039~\cite{we}. However, the full post-Keplerian test is necessary because the masses of binary systems components predicted by the theory can differ from their values in GR, which can affect the final restrictions imposed on the model~\cite{stars}.
	
It would also be great to get the universal apparatus for testing gravitational theories with massive fields in the weak-field limit as the original PPN formalism for massless, but this is a topic of more extensive research in the future.

The authors thank N. A. Avdeev and V.V. Kolybasova  for discussions and comments on the topics of this paper. This work was supported by the grant 18-32-00785 from Russian Foundation for Basic Research.

	\appendix
	\section{Point-mass and perfect fluid PPN metric}\label{app1}
	Point-mass metric~\cite{ppn1}:
	\begin{wide}
	\begin{eqnarray}\label{pmmet}
	g_{00}=&-&1+2\sum_{k}\frac{G}{c^2}\frac{m_k}{r_k}-2\beta\left(\sum_{k}\frac{G}{c^2}\frac{m_k}{r_k}\right)^2+2(1-2\beta+\zeta_2)\sum_{k}\frac{G}{c^2}\frac{m_k}{r_k}\sum_{j\neq k}\frac{G}{c^2}\frac{m_j}{r_{jk}}+(2\gamma+1+\alpha_3+\zeta_1)\sum_{k}\frac{G}{c^4}\frac{m_kv_k^2}{r_k}\nonumber
	\\
	&-&\zeta_1\sum_{k}\frac{G}{c^4}\frac{m_k}{r_k^3}( \mathbf{v}_k   \mathbf{r}_k)^2-(\alpha_1-\alpha_2-\alpha_3)w^2\sum_{k}\frac{G}{c^4}\frac{m_k}{r_k^3}-\alpha_2\sum_{k}\frac{G}{c^4}\frac{m_k}{r_k}( \mathbf{w}   \mathbf{r}_k)^2+(2\alpha_3-\alpha_1)\sum_{k}\frac{G}{c^4}\frac{m_k}{r_k}( \mathbf{w}   \mathbf{v}_k),\nonumber\\
	g_{0j}=&-&\frac{1}{2}(4\gamma+3+\alpha_1-\alpha_2+\zeta_1)\sum_{k}\frac{G}{c^3}\frac{m_kv_k^j}{r_k}-\frac{1}{2}(1+\alpha_2-\zeta_1)\sum_{k}\frac{G}{c^3}\frac{m_k}{r_k^3}( \mathbf{v}_k   \mathbf{r}_k)r_k^j-\frac{1}{2}(\alpha_1-2\alpha_2)w^j\sum_{k}\frac{G}{c^3}\frac{m_k}{r_k}\nonumber\\
	&+&\alpha_2\sum_{k}\frac{G}{c^3}\frac{m_k}{r_k^3}( \mathbf{w}  \mathbf{ r}_k)r_k^j,\nonumber\\
	g_{ij}= &&\left(1+2\gamma\sum_{k}\frac{G}{c^2}\frac{m_k}{r_k}\right) \delta_{ij},
	 \end{eqnarray}
	 \end{wide}
	here $w^i$ is coordinate velocity of PPN coordinate system relative to the mean rest frame of the universe.
	
	Perfect fluid metric~\cite{Will}:
\begin{wide}
	\begin{eqnarray}\label{pfmet}
	g_{00}=&-&1+2\frac{1}{c^2}U-2\beta\frac{1}{c^4} U^2+(2\gamma+1+\alpha_3+\zeta_1-2\xi)\frac{1}{c^4}\Phi_1-2(2\beta-1-\zeta_2-\xi)\frac{1}{c^4}\Phi_2+2(1+\zeta_3)\frac{1}{c^4}\Phi_3+\frac{1}{c^4}\Phi^{PF}\nonumber\\
	&+&2(3\gamma+3\zeta_4-2\xi)\frac{1}{c^4}\Phi_4-(\zeta_1-2\xi)\frac{1}{c^4}\Phi_6-2\xi\frac{1}{c^4}\Phi_W,\nonumber\\
	g_{0j}=&-&\frac{1}{2}[4\gamma+3+\alpha_1-\alpha_2+\zeta_1-2\xi]\frac{1}{c^3}V_j-\frac{1}{2}(1+\alpha_2-\zeta_1+2\xi)\frac{1}{c^3}W_j +\frac{1}{c^3}\Phi_j^{PF},\nonumber\\
	g_{ij}=&&\left(1+2\gamma\frac{1}{c^2}U\right) \delta_{ij},
	 \end{eqnarray}
	 \end{wide}
	where PPN potentials are represented
	\begin{wide}
	\begin{eqnarray}\label{potentials}
		U&=&\int G\frac{\rho'}{| \mathbf{r}- \mathbf{r'}|}d^3 \mathbf{r}',\quad\Phi_1=\int G\frac{\rho'v'^2}{| \mathbf{r}- \mathbf{r'}|}d^3 \mathbf{r}',\quad\Phi_2=\int G\frac{\rho'U'}{| \mathbf{r}- \mathbf{r'}|}d^3 \mathbf{r}', \quad\Phi_3=\int G\frac{\rho'\Pi'}{| \mathbf{r}- \mathbf{r'}|}d^3 \mathbf{r}',	\nonumber\\
		\Phi_4&=&\int G\frac{p'}{| \mathbf{r}- \mathbf{r'}|}d^3 \mathbf{r}',\quad V_j=\int G\frac{\rho v_j}{| \mathbf{r}- \mathbf{r'}|}d^3 \mathbf{r}',\quad\Phi_6=\int G\rho'v_j'v_k'\frac{(r-r')^j(r-r')^k}{| \mathbf{r}- \mathbf{r'}|^3}d^3 \mathbf{r}',\nonumber\\
		 \Phi_j^{PF}&=&-\frac{1}{2}\alpha_1w_jU+\alpha_2w^iU_{ij},\Phi_W=\int G^2\rho'\rho''\frac{(r-r')_j}{| \mathbf{r}- \mathbf{r'}|^3}\left[\frac{(r'-r'')^j}{| \mathbf{r}- \mathbf{r''}|}-\frac{(r-r'')^j}{| \mathbf{r'}- \mathbf{r''}|}\right]d^3 \mathbf{r}'d^3 \mathbf{r}'',\nonumber\\
W_j&=&\int G\frac{\rho' \mathbf{v}'( \mathbf{r}- \mathbf{r}')(r-r')_j}{| \mathbf{r}- \mathbf{r'}|^3}d^3 \mathbf{r}', 	U_{ij}=\int G\frac{\rho'(r-r')_i(r-r')_j}{| \mathbf{r}- \mathbf{r'}|}d^3 \mathbf{r}',\nonumber\\
	\Phi^{PF}&=&(\alpha_3-\alpha_1)w^2U+\alpha_2w^iw^jU_{ij}+(2\alpha_3-\alpha_1)w^jV_j,
	 \end{eqnarray}
	 \end{wide}
	here ''PF''-potentials are responsible for preferred frames effects.
	
		\begin{wide}
	\section{The perfect fluid metric of the hybrid f(R)-gravity}\label{app2}
	The perfect fluid metric of hybrid f(R)-gravity:

		\begin{eqnarray}\label{metr}
 	g_{00}=&-&1+\frac{k^2}{4\pi(1+\phi_0)c^2} \int \frac{\rho'}{| \mathbf{r}- \mathbf{r}'|}\left(1-\frac{\phi_0}{3}   e^{-m_\varphi | \mathbf{r}- \mathbf{r}'|}\right)d^3 \mathbf{r}'-\frac{k^4}{32\pi^2(1+\phi_0)^2c^4}   \nonumber\\
 	&\times&\int \frac{\rho'}{| \mathbf{r}- \mathbf{r}'|}\left(1-\frac{\phi_0}{3}   e^{-m_\varphi | \mathbf{r}- \mathbf{r}'|}\right)\frac{\rho''}{| \mathbf{r}- \mathbf{r}''|}\left(1-\frac{\phi_0}{3}   e^{-m_\varphi | \mathbf{r}- \mathbf{r}''|}\right)d^3 \mathbf{r}''d^3 \mathbf{r}'+\frac{k^4}{32\pi^2(1+\phi_0)^2c^4}\frac{\phi_0(1+\phi_0)}{18}\nonumber\\
 	&\times&\int \frac{\rho'}{| \mathbf{r}- \mathbf{r}'|}  e^{-m_\varphi| \mathbf{r}- \mathbf{r}'|}\frac{\rho''}{| \mathbf{r}- \mathbf{r}''|}  e^{-m_\varphi| \mathbf{r}- \mathbf{r}''|}d^3 \mathbf{r}''d^3 \mathbf{r}'+\frac{k^2}{4\pi(1+\phi_0)c^4} \int\frac{\Pi'\rho'}{| \mathbf{r}- \mathbf{r}'|}  \biggl(1-\frac{\phi_0}{3}   e^{-m_\varphi| \mathbf{r}- \mathbf{r}'|}  \biggr)d^3 \mathbf{r}'\nonumber\\
	 &+&\frac{k^2}{2\pi(1+\phi_0)c^4}\int\frac{\rho'v'^2}{| \mathbf{r}- \mathbf{r}'|}d^3 \mathbf{r}'+\frac{3k^2}{4\pi(1+\phi_0)c^4}\int \frac{p'}{| \mathbf{r}- \mathbf{r}'|}  \biggl(1+\frac{\phi_0}{3}   e^{-m_\varphi| \mathbf{r}- \mathbf{r}'|}  \biggr)d^3 \mathbf{r}'\nonumber\\
	 &+&\frac{k^4}{32\pi^2(1+\phi_0)^2c^4}\int \frac{\rho'}{| \mathbf{r}- \mathbf{r}'|}\left[1-\frac{\phi_0}{3}   e^{-m_\varphi| \mathbf{r}- \mathbf{r}'|}\right]d^3 \mathbf{r}'\int\frac{\hat\rho}{| \mathbf{r'}- \mathbf{\hat r}|} \biggl(1-\frac{\phi_0}{3}   e^{-m_\varphi | \mathbf{r'}- \mathbf{\hat r}|}  \biggr)d^3  \mathbf{\hat r}\nonumber\\
&+&\frac{3k^4}{32\pi^2(1+\phi_0)^2c^4} \int \frac{\rho'}{| \mathbf{r}- \mathbf{r}'|}\left[1-\frac{\phi_0}{3}   e^{-m_\varphi| \mathbf{r}- \mathbf{r}'|}\right]d^3 \mathbf{r}'\int\frac{\hat\rho}{| \mathbf{r'}-\mathbf{\hat r}|} \biggl(1+\frac{\phi_0}{3}   e^{-m_\varphi | \mathbf{r'}-\mathbf{\hat r}|}  \biggr)d^3\mathbf{\hat r}\nonumber\\
&-&\frac{k^4}{16\pi^2(1+\phi_0)^2c^4} \int \frac{\rho'}{| \mathbf{r}- \mathbf{r}'|}\left[1-\frac{\phi_0}{3}   e^{-m_\varphi| \mathbf{r}- \mathbf{r}'|}\right]d^3 \mathbf{r}'\int\frac{\hat\rho}{| \mathbf{r'}- \mathbf{\hat r}|}  \biggl(1-\frac{\phi_0}{3}   e^{-m_\varphi | \mathbf{r'}- \mathbf{\hat r}|}  \biggr)d^3  \mathbf{\hat r}\nonumber\\
 &+&\frac{k^4}{16\pi^2(1+\phi_0)^2c^4}\frac{\phi_0(1+\phi_0)}{18}\int \frac{\rho'}{| \mathbf{r}- \mathbf{r}'|}  e^{-m_\varphi| \mathbf{r}- \mathbf{r}'|}d^3 \mathbf{r}'\int\frac{\hat\rho}{| \mathbf{r'}- \mathbf{\hat r}|} e^{-m_\varphi| \mathbf{r'}- \mathbf{\hat r}|}d^3  \mathbf{\hat r}\nonumber\\
&+&\frac{(7\phi_0+1)k^4\phi_0}{2304\pi^3(1+\phi_0)^2c^4}  m_\varphi^2\int\frac{ e^{-m_\varphi| \mathbf{r}- \mathbf{r}'|}}{| \mathbf{r}- \mathbf{r}'|}d^3 \mathbf{r}'  \biggl(\int \frac{\hat\rho}{| \mathbf{r'}- \mathbf{\hat r}|}   e^{-m_\varphi | \mathbf{r'}-\mathbf{\hat r}|}\frac{\rho''}{| \mathbf{r'}- \mathbf{r''}|}   e^{-m_\varphi | \mathbf{r'}- \mathbf {r''}|}d^3\mathbf{\hat r}d^3 \mathbf{ r''}  \biggr)\nonumber\\
&+&\frac{k^4\phi_0}{192\pi^3(1+\phi_0)^2c^4}  m_\varphi\int\frac{ e^{-m_\varphi| \mathbf{r}- \mathbf{r}'|}}{| \mathbf{r}- \mathbf{r}'|}d^3 \mathbf{r}'  \biggl(\int \frac{\hat\rho}{| \mathbf{r'}-\mathbf{\hat r}|}   e^{-m_\varphi | \mathbf{r'}-\mathbf{\hat r}|}\frac{\rho''}{| \mathbf{r'}- \mathbf{r''}|}   e^{-m_\varphi | \mathbf{r'}- \mathbf {r''}|}d^3\mathbf{\hat r}d^3 \mathbf{ r''}  \biggr)\nonumber\\
&+&\frac{k^4\phi_0^2}{1728\pi^3c^4}   \big[V''-\frac{\phi_0}{2}V'''  \big]\int\frac{ e^{-m_\varphi| \mathbf{r}- \mathbf{r}'|}}{| \mathbf{r}- \mathbf{r}'|}d^3 \mathbf{r}'  \biggl(\int \frac{\hat\rho}{| \mathbf{r'}-\mathbf{\hat r}|}   e^{-m_\varphi | \mathbf{r'}-\mathbf{\hat r}|}\frac{\rho''}{| \mathbf{r'}- \mathbf{r''}|}   e^{-m_\varphi | \mathbf{r'}- \mathbf {r''}|}d^3\mathbf{\hat r}d^3 \mathbf{ r''}  \biggr),\nonumber \\
 	g_{0i}=&-&\frac{k^2}{4\pi(1+\phi_0)c^3}\int\frac{\rho'v_i'}{| \mathbf{r}- \mathbf{r'}|}  \biggl(1+\frac{\phi_0}{3} e^{-m_\varphi| \mathbf{r}- \mathbf{r}'|}  \biggr)d^3 \mathbf{r}'-\frac{3k^2}{16\pi(1+\phi_0)c^3}\int\frac{\rho'v_i'}{| \mathbf{r}- \mathbf{r'}|}  \biggl(1-\frac{\phi_0}{3} e^{-m_\varphi| \mathbf{r}- \mathbf{r}'|}  \biggr)d^3 \mathbf{r}'\nonumber\\
&-&\frac{k^2}{16\pi(1+\phi_0)c^3}\int\frac{\rho'x_i'( \mathbf{v'}\cdot \mathbf{r'})}{| \mathbf{r}- \mathbf{r'}|^3}\left[1-\frac{\phi_0}{3} e^{-m_\varphi| \mathbf{r}- \mathbf{r}'|}\right]d^3 \mathbf{r},\nonumber\\
 	g_{ij}= &&\delta_{ij}  \biggl(1+\frac{k^2}{4\pi(1+\phi_0)c^2}  \int \frac{\rho'}{| \mathbf{r}- \mathbf{r}'|}\left(1+\frac{\phi_0}{3}   e^{-m_\varphi| \mathbf{r}- \mathbf{r}'|}\right)d^3 \mathbf{r}'  \biggr).
	 \end{eqnarray}
	
\end{wide}


\clearpage
\begin{references}
\bibitem{DE1}
		S.~Perlmutter et al., ApJ {\bf 517},  565 (1999).
		
		\bibitem{DE2}
		A.~G.~Riess et al.,  AJ {\bf 116}, 1009 	(1998).
		
		\bibitem{DE3}
		A.~G.~Riess et al.,  ApJ {\bf 607}, 665 	(2004).
		
		\bibitem{DM1}
		F.~Zwicky, Helvetica Physica Acta {\bf 6}, 110 (1933).
		
		\bibitem{DM2}
		  J.~H.~Oort, Bulletin of the Astronomical Institutes of the Netherlands {\bf 6}, 249 (1932).		  
		 
		 \bibitem{fr1}
		P.~G.~Bergmann, International Journal of Theoretical Physics {\bf 1},  25 (1968).
		
		\bibitem{fr2}
		A.~De Felice and S.~Tsujikawa, Living Reviews in Relativity {\bf 13},  3 (2010).
		  
		\bibitem{odintsov9}
		S.~Nojiri, S.~D.~Odintsov and V.~K.~Oikonomou, Phys.Rept. {\bf 692}, 1 (2017).
		
		\bibitem{odintsov10}
		S.~Nojiri and S.~D.~Odintsov, Phys. Rept. {\bf505}, 59 (2011).
		
		\bibitem{starobinsky}
		A.~A.~Starobinsky, Phys. Lett. B {\bf 91},  99 (1980).
		
		
		\bibitem{odintsov1}
		S.~Nojiri and S.~D.~Odintsov, Phys. Rev. D {\bf 68}, 123512 (2003).
		
		\bibitem{odintsov2}
		  F.~Briscese, E.~Elizalde, S.~Nojiri and S.~D.~Odintsov, Phys. Lett. B {\bf 646}, 105 (2007).
		  
		  \bibitem{saez}
		   D.~Saez-Gomez, Gen. Rel. Grav. {\bf 41}, 1527 (2009). 
		   
		   \bibitem{odintsov3}
		   S.~Nojiri and S.~D.~Odintsov, Phys. Lett. B {\bf 657}, 238 (2007).
		   
		   \bibitem{odintsov4}
		   S.~Nojiri and S.~D.~Odintsov, Phys. Rev. D {\bf 77}, 026007 (2008). 
		   
		   
		   \bibitem{odintsov5}
		    S.~Nojiri, S.~D.~Odintsov and D.~Saez-Gomez, Phys. Lett. B {\bf 681}, 74 (2009).
		    
		  \bibitem{odintsov6}
		   G.~Cognola, E.~Elizalde, S.~D.~Odintsov, P.~Tretyakov and S.~Zerbini, Phys. Rev. D {\bf 79}, 044001 (2009).
		   
		   \bibitem{odintsov7}
		   G.~Cognola, E.~Elizalde, S.~Nojiri, S.~D.~Odintsov, L.~Sebastiani and S.~Zerbini, Phys. Rev. D {\bf 77}, 046009 (2008). 
		   
		   \bibitem{odintsov8}
		   S.~D.~Odintsov, D.~Saez-Gomez and G.~S.~Sharov,  Eur.Phys.J. C {\bf 77}, 862 (2017).
		   
		   \bibitem{capo1}
		S.~Capozziello and M.~Francaviglia, Gen. Rel. Grav. {\bf 40},  357 (2008).
		
		\bibitem{capo2}
		 T.~P.~Sotiriou and V.~Faraoni, Rev. Mod. Phys. {\bf 82}, 451 (2010).
		 
		 	\bibitem{chiba}
		T.~Chiba, Phys. Lett. B {\bf 575},  1 (2003).
		
		\bibitem{olmo1}
		 G.~J.~Olmo, Phys. Rev. Lett. {\bf 95}, 261102 (2005).
		
		\bibitem{olmo2}
		G.~J.~Olmo, Phys. Rev. D {\bf 75}, 023511 (2007).
		 
		  \bibitem{khoury}
		  J.~Khoury and A.~Weltman, Phys. Rev. Lett. {\bf 93}, 171104 (2004).
		  
		  \bibitem{capo0}
		   S.~Capozziello and S.~Tsujikawa, Phys.Rev. D {\bf 77}, 107501 (2008).
		   
		   \bibitem{khoury1}
		J.~Khoury and A.~Weltman, Phys. Rev. D {\bf69}, 044026 (2004).
		
		\bibitem{hu}
		 W.~Hu and I.~Sawicki, Phys. Rev. D {\bf76}, 064004 (2007) .
		   
		   	\bibitem{koivisto1}
		 T.~Koivisto and H.~Kurki-Suonio, Class. Quantum Grav. {\bf 23},  2355 (2006).
		
		\bibitem{koivisto2}
		 T.~Koivisto, Phys. Rev. D {\bf 73}, 083517 (2006).
		
		\bibitem{hybrid1}
		T.~Harko, T.~S.~Koivisto, F.~S.~N.~Lobo and G.~J.~Olmo,  Phys. Rev. D {\bf 85}, 084016 (2012).
		
		 \bibitem{bohmer}
		  C.~G.~B\"ohmer, F.~S.~N.~Lobo and N.~Tamanini, Phys. Rev. D {\bf 88}, 104019 (2013).
		 
		 \bibitem{lima}
		 N.~A.~Lima and V.~Smer-Barreto, ApJ {\bf818}, 186 (2016).
		 
		 \bibitem{lea} 
		I.~Leanizbarrutia, F.~S.~N.~Lobo and D.~S\'aez-G\'omez, Phys. Rev. D {\bf 95}, 084046 (2017).
		 
		 \bibitem{capo3}
		 S.~Capozziello, T.~Harko, T.~S.~Koivisto, F.~S.~N.~Lobo and G.~J.~Olmo, JCAP {\bf 04}, 011 (2013).
		 
		 \bibitem{capo4}
		  S.~Capozziello, T.~Harko, T.~S.~Koivisto, F.~S.~N.~Lobo and G.~J.~Olmo, JCAP {\bf07}, 024 (2013).
		  
		  \bibitem{capo5}
		 S.~Capozziello, T.~Harko,  T.~S.~Koivisto, F.~S.~N.~Lobo and G.~J.~Olmo, Astroparticle Physics {\bf50-52C}, 65 (2013).
		 
		 \bibitem{capo6}
		 S.~Capozziello, T.~Harko,  T.~S.~Koivisto, F.~S.~N.~Lobo and G.~J.~Olmo, Phys. Rev. D {\bf86}, 127504 (2012).
		 
		 \bibitem{stars}
		B.~Danila, T.~Harko,   F.~S.~N.~Lobo and M.~K.~Mak, Phys. Rev. D {\bf 95}, 044031 (2017).
		
		\bibitem{hybrid2}
		S.~Capozziello,  T.~Harko, T.~S.~Koivisto, F.~S.~N.~Lobo and G.~J.~Olmo,  Universe {\bf 1}, 199 (2015).
		
		
		\bibitem{mercury1}
		C.~M.~Will, Phys. Rev. Lett. {\bf 120}, 191101  (2018).
		
		\bibitem{mercury2}
		A.~Fienga, J.~Laskar, P.~Kuchynka,  H.~Manche, G.~Desvignes, M.~Gastineau,  I.~Cognard and G.~Theureau, Celestial Mechanics and Dynamical Astronomy {\bf 111}, 363 (2011).

		\bibitem{mercury3}
		A.~Verma, A.~Fienga, J.~Laskar, H.~Manche and M.~Gastineau,  Astronomy \& Astrophysics {\bf 561},  A115 (2014).

		\bibitem{mercury4}
		A.~Fienga,  J.~Laskar, P.~Exertier, H.~Manche and M.~Gastineau, Celestial Mechanics and Dynamical Astronomy {\bf 123}, 325	(2015).
		
		\bibitem{PPN1}
		K.~Nordtvedt, Phys. Rev. {\bf 169}, 1017 (1968).
		
		\bibitem{PPN2}
		C.~M.~Will, ApJ {\bf 163}, 611 (1971).
		
		\bibitem{ppn1}
		 C.~M.~Will and  K.~Nordtvedt K, ApJ. {\bf 177}, 757 (1972).
		
		\bibitem{will1}
		C.~M.~Will, {\it Theory and Experiment in Gravitational Physics}, Cambridge University Press, Cambridge, UK (1993).
		
		\bibitem{massbd}
		J.~Alsing, E.~Berti, C.~M.~Will and H.~Zaglauer, Phys. Rev. D {\bf 85}, 064041 (2012).
		
		\bibitem{helbig}
		T.~Helbig, ApJ {\bf 382},  223 (1991).
		
		\bibitem{massivebd}
		L.~Perivolaropoulos, Phys. Rev. D {\bf 81},  047501  (2010).
		
			\bibitem{Nutku}
		 Y.~Nutku, ApJ {\bf 155}, 999 (1969).		
						
		
	

	
		\bibitem{gamma}
		B.~Bertotti, L.~Iess and P.~Tortora, Nature {\bf 425} 374 (2003).

		
		\bibitem{will14}
		 C.~M.~Will, Living Reviews in Relativity {\bf 17}, 4 (2014).
		
		
		
		\bibitem{satirou}
		T.~P.~Sotiriou and E.~Barausse, Phys. Rev. D {\bf 75}, 084007 (2007)
		
		\bibitem{lee}
		D.~L.~Lee,  Phys. Rev. D {\bf 10},  2374 (1974).
		
		\bibitem{Will}
		E.~Poisson and C.~M.~Will, {\it Gravity: Newtonian, Post-Newtonian, Relativistic}, Cambridge University Press, Cambridge, UK (2014).

		\bibitem{newmes}
		R.~S.~Park, W.~M.~Folkner, A.~S.~Konopliv, J.~G.~Williams, D.~E.~Smith and  M.~T.~Zuber, AJ {\bf153}, 3, 121 (2017).
		
	
		\bibitem{geff}
		P.~J.~Mohr, D.~B.~Newell and  B.~N.~Taylor, Rev. Mod. Phys. {\bf 88},  035009 (2016).

		\bibitem{we}
		P.~I.~Dyadina, N.~A.~Avdeev and S.~O.~Alexeyev, Monthly Notices of the Royal Astronomical Society {\bf 483}, 947	(2019).
				
	


\end{references}
\end{document}